# Wearables and location tracking technologies for mental-state sensing in outdoor environments

**Amit Birenboim[1,2], Martin Dijst[1,3], Floortje E. Scheepers[4], Maartje Poelman[1] and Marco Helbich[1]**

[1] Department of Human Geography and Spatial Planning, Utrecht University, Princetonlaan 8a, 3584 CB, Utrecht, The Netherlands
[2] Department of Geography and the Human Environment, Tel Aviv University, Zelig 10, 6997801, Tel Aviv, Israel
[3] Department of Urban Development and Mobility, Luxembourg Institute of Socio-Economic Research, Porte des Sciences 11, 4366, Esch-sur-Alzette, Luxembourg
[4] Department of Psychiatry, University Medical Center, Heidelberglaan 100, 3508 GA, Utrecht The Netherlands

\* Corresponding author, Amit Birenboim: abirenboim@tauex.tau.ac.il; Tel.: +972-3-6406243

# Abstract

Advances in commercial wearable devices are increasingly facilitating the collection and analysis of everyday physiological data. This paper discusses the theoretical and practical aspects of using such ambulatory devices for the detection of episodic changes in physiological signals as a marker for mental state in outdoor environments. **A** pilot study was conducted to evaluate the feasibility of utilizing commercial wearables in combination with location tracking technologies. The study measured physiological signals for 15 participants, including heart rate, heart-rate variability, and skin conductance. Participants' signals were recorded during an outdoor walk that was tracked using a GPS logger. The walk was designed to pass through various types of environments including green, blue, and urban spaces as well as a more stressful road crossing. The data that was obtained was used to demonstrate how biosensors information can be contextualized and enriched using location information. Significant episodic changes in physiological signals under real-world conditions were detectable in the stressful road crossing, but not in the other types of environments. The article concludes that despite challenges and limitations of current off-the-shelf wearables, the utilization of these devices offers novel opportunities for evaluating episodic changes in physiological signals as a marker for mental state during everyday activities including in outdoor environments. ***Keywords:*** *Wearable, Mental state, GPS, Stress, Electrodermal activity.*

Recent years have seen a steady increase in mental health disorders worldwide (Whiteford et al. 2013). Interestingly, psychiatric disorders including schizophrenia, mood and anxiety disorders are commonly more prevalent in urban environments (Peen et al. 2010; Lederbogen et al. 2011). This emphasizes the relevancy of environmental exposures to understanding mental state and raises questions as to the mechanisms through which environments affect such health outcomes. Environmental psychologists have already identified the beneficial influence that natural green environments have on stress reduction and attention restoration few decades ago (Ulrich 1984; Kaplan and Kaplan 1989). More recently, the discourse regarding therapeutic and adverse environments and landscapes has been promoted in the geographical, urban and health literature (Evans 2003; Gong et al. 2016). In this respect, the effect of greenery on mental health outcomes is probably the single most intensively researched environmental quality (Bowler et al. 2010; Helbich et al. 2018). Other environmental elements that were studied include blue spaces such as canals and seashores (Wheeler et al. 2012), traffic load (Healey and Picard 2005), social environmental characteristics (Lorenc et al. 2012) and more.

With some exceptions, the investigation of the association between the environment and mental state relied mainly on aggregative and static environmental factors (i.e. cross-sectional city or neighborhood characteristics). However, there is a growing agreement in recent years that in order to better understand what are the exact environmental elements and actual mechanisms through which the environment affects mental state, a more dynamic investigation is required (Chaix 2018; Helbich 2018). As a result, researchers have been looking for tools that will allow a closer and more objective examination of the moment-by-moment environmental exposure and its impact on health outcomes. A main facilitator of this trend is the introduction of new sensing capabilities of both external (mainly the physical environment) and internal (i.e. the personal context) which are becoming more prevalent, especially in urban environments (Sagl, Resch, and Blaschke 2015). In particular, location tracking technologies most notably the GPS allow collecting high tempo-spatial resolution information about individuals location and hence their environmental exposure and obtain additional contextual information (Chaix 2018)

With the introduction of new wearable biosensors in the market, efforts are being invested in applying the continuous stream of physiological data supplied by these devices to basic research, clinical applications, and practices of "quantified self" (Swan 2013; Reeder and David 2016; Li et al. 2017; Wright et al. 2017). However, ambulatory, real-world measurements of physiological signals through non-invasive wearables pose several methodological challenges for researchers, especially in cases in which signals are used as markers for mental states. First, in most—if not all—cases, measurement quality of ambulatory devices is inferior to that of lab instruments due to technical constraints of battery life and physical dimensions. For example, wearables are often equipped with inferior technology such as photoplethysmography (a low-cost non-invasive optical technology in which skin light absorption is measured to evaluate various cardiovascular indicators) rather than the more reliable electrocardiography that is commonly used in hospitals (Lin et al. 2014) and worn in suboptimal locations such as the wrist (van Dooren, de Vries, and Janssen 2012). Second, researchers using ambulatory devices have less control over the environmental factors and stimuli that their subjects are exposed to than do those conducting experiments under laboratory conditions. This in turn makes it difficult to isolate the impact of specific stimuli. These types of deficiencies, typical to most field research, weaken the internal validity of results (Wilhelm and Grossman 2010). Third, and related to the previous drawback, real-world measurements of physiological signals—especially those conducted out of doors—are often fraught with noise and measurement errors, making data interpretation even more demanding (Sun et al. 2010; Osborne and Jones 2017). For example, skin conductivity which rises during emotional arousal will also increase as a result of extraneous variables such as high ambient temperature that increases sweating. In addition, measurement errors are more common out-of-the-lab and especially when participants are engaged in physical activity (e.g. walking) that interrupts the smooth functioning of the wearables.

Despite these challenges, researchers have shown increased interest in ambulatory measurements of physiological signals to detect changes in stress and other mental states during everyday activities (Hartig et al. 2003; Healey and Picard 2005; Wilhelm, Pfaltz, and Grossman 2006; Bakker, Pechenizkiy, and Sidorova 2011; Sharma and Gedeon 2012; de Faria, da Silva, and Cugnasca 2016; Osborne and Jones 2017). While a key catalyst for the growing interest in this type of measurement is the development of new wearable biosensors that can be conveniently utilized in daily life (Wright et al. 2017), an additional factor has promoted the use of wearables in research: the increased focus on ecological approaches in behavior and health research in the last decade or so (McLaren and Hawe 2004; Fahrenberg et al. 2007). These approaches call into question researchers' ability to explain emotional functioning in real life based on laboratory studies alone (Wilhelm and Grossman 2010), and therefore facilitate the development of tools that can supply reliable information about mental states in naturally occurring environments (Fahrenberg et al. 2007; Eskes et al. 2016; Birenboim 2018).

In this regard, wearable biosensors have at least four major advantages over traditional data-collection methods such as surveys, questionnaires, and one-time measurements of physiological signals:

- Real-time physiological signals recorded by wearable sensors are considered more objective than are self-reported assessments, which tend to be biased (Wilhelm and Grossman 2010; Sharma and Gedeon 2012).
- Wearables allow for continuous measurement at a high temporal resolution of parts of seconds (Healey and Picard 2005). This resolution cannot be obtained when relying on one-time measurements or self-report surveys alone.
- Wearables significantly reduce the burden on participants, who are not required to repeatedly complete surveys. This makes an extended data-collection period—ranging from a few hours to several months—possible.
- Finally, and of key importance to ecological approaches, ambulatory measurements facilitate the investigation of people's physiological signals during their daily routines in real-life situations, offering greater ecological validity than do lab studies (Wilhelm and Grossman 2010).

However, with a few exceptions (see for example: Sun et al. 2010; Osborne and Jones 2017; Shoval, Schvimer, and Tamir 2018a, 2018b), real-world measurements of physiological signals have by and large been restricted to static postures such as a sedentary driving position (Healey, Seger, and Picard 1999; Healey and Picard 2005) and/or to studies that focus on long-term behavioral trends (i.e., hourly/daily changes) rather than second-by-second physiological reactions (Wilhelm and Grossman 2010, 553). There are two primary reasons for this. The first is that the quality of physiological data is significantly reduced when measuring subjects who move, since, as mentioned above, movement increases measurement errors on the part of the sensors. The second is that the social and physical context, which is essential for interpreting the results (Bakker, Pechenizkiy, and Sidorova 2011; Osborne and Jones 2017), changes frequently when people move. In order to deal with this problem, contextual information needs to be collected continuously (Sun et al. 2010), complicating data collection and research design. Contextual data may include, for example, information about the surrounding environment, type and intensity of activity, and social context (e.g., stressful job interview vs. enjoyable social event).

Emerging sensing technology that makes possible convenient daily measurements of physiological signals in real life has significant clinical, research, and commercial potential (Blaauw et al. 2016). It can be used to detect changes in stress levels throughout the day (Bakker, Pechenizkiy, and Sidorova 2011), to study the association between environment and momentary mental well-being (Hartig et al. 2003), to serve as a diagnostic and intervention tool for psychiatric problems (MacLean, Roseway, and Czerwinski 2013), to enhance practices of quantified self (Shin and Biocca 2017) and support processes of urban planning and management (Resch et al. 2015; Sagl, Resch, and Blaschke 2015). Here we suggest augmenting bio-sensors data with spatial information generated by location tracking technologies (e.g. GPS). The high tempo-spatial resolution of location information that current technologies generate in combination with geographical layers and other external sources of information allow augmenting bio-sensors information with contextual information about the surrounding environment and about the activity one is engaged with. This may include additional information about land use, buildings density, weather and movement parameters such as speed. All of which could be essential for utilizing bio-sensors data as a marker for mental state.

Given both the potential and the challenges that come with emerging sensing technology, this article examines the adequacy of utilizing current off-the-shelf wearables in combination with location tracking technologies to serve as a marker for mental state in outdoor environments in high temporal resolution, and especially during walks in urban landscapes. For this objective, we tested the functionality of two off-the-shelf wearables, the Empatica E4 wristband and Microsoft Band 2 (henceforth "MS Band") in combination with a GPS information during a controlled outdoor walk in an urban setting. This technique may allow a close investigation of the impact of environmental factors (e.g. green spaces) on our daily wellbeing.

**Physiological Signals as Markers for Mental States**

The most commonly used physiological signals for inferring changes in mental state are those associated with the activity of the autonomic nervous system (Kreibig et al. 2007). The autonomic nervous system—with its two branches, the sympathetic and parasympathetic nervous systems—acts largely unconsciously, taking part in the regulation of bodily functions such as the activity of the heart and lungs, digestion, pupillary response, and sexual arousal. It is thought to play a major role in the fight-or-flight response during events that are conceived to pose a threat to one's survival (Cannon 1929; Kreibig et al. 2007). Emotional reactions (Kreibig 2010) such as psychological stress (Jansen et al. 1995) seem to correspond with this fight-or-flight response.

Physiological signals from the autonomic nervous system can be extracted from different bodily systems or organs to make inferences about an individual's mental state. These include, but are not limited to, the cardiovascular system (Appelhans and Luecken 2006), skin (Rimm-Kaufman and Kagan 1996; Boucsein et al. 2012), respiratory system (Boiten 1998), endocrine system (Almeida, McGonagle, and King 2009), and eyes (Bradley et al. 2008). Due to the ease of recording them outside the lab, the physiological signals of the cardiovascular system and skin are most commonly recorded in research using ambulatory measurements. Thus in this study we analyzed the following physiological signals from these two systems using the E4 and the MS Band which were utilized in other studies to record physiological signals (Lopez-Samaniego and Garcia-Zapirain 2016; Osborne and Jones 2017):

- Heart rate (HR): This measure, often represented by the number of heart beats per minute, is the most commonly utilized physiological signal for monitoring changes in mental state. During a fight-or-flight response, the sympathetic system increases heart activity, allowing the body to respond more efficiently to external threats. Increased HR is associated with stress (Taelman et al. 2009) and emotions of anger, anxiety, embarrassment, fear, happiness, joy, and surprise. In contrast, lower HR levels are associated with a state of serenity and emotions such as acute sadness, affection, and contentment (Kreibig 2010).

- Heart rate variability (HRV): HRV takes into account the variation between the heart's beat-to-beat intervals, also known as interbeat intervals. A stimulated sympathetic system results in lower HRV levels. In contrast, when an individual is relaxed, the tone of the parasympathetic system increases; this, in turn, results in a greater interbeat interval variation (Appelhans and Luecken 2006). There are several indicators that assess HRV (Appelhans and Luecken 2006; Kreibig 2010). In this study, we calculated three common indexes using Kubious HRV 2.2 software (Tarvainen et al. 2014):
    - SDNN: the standard deviation of interbeat intervals within a given time window.
    - pNN50: the ratio between the number of successive pairs of interbeat intervals that differ in more than 50 milliseconds from one another and the total number of interbeat intervals within a time window.

- LF/HF: a frequency domain measurement that divides the variance of continuous interbeat interval series into its frequency components. The low frequency (LF) band is typically set to 0.04–0.15 Hz and represents the activity of both the sympathetic and parasympathetic systems. The high frequency (HF), which is typically set to 0.15–0.4 Hz, represents the activity of the parasympathetic system alone. Thus, the greater the LF/HF ratio, the greater the tone of the sympathetic system is.

HRV indexes can be extracted for both the long (e.g., daily) and short (e.g., five-minute) term. While a low HRV is associated with psychological stress (Appelhans and Luecken 2006) and with emotions such as anger, anxiety, fear, and happiness, high levels of HRV correlate with more relaxed states, but also with a sense of amusement (Kreibig 2010).

- Electrodermal activity (EDA): also known as Galvanic Skin Response, this refers to the variation in the electrical properties of the skin (i.e., skin conductance/resistance). EDA is regulated by the sympathetic nervous system through the sweat glands. When stimulated (e.g., due to emotional arousal), the sympathetic nervous system will intensify sweating, which in turn will increase skin conductivity. High EDA levels are associated with psychological stress (Healey and Picard 2005) and feelings of anger, anxiety, fear, and amusement. Lower EDA levels correlate with more relaxed states and with acute sadness and a sense of relief (Kreibig 2010). Raw EDA data is typically divided into two components: (1) the skin conductance level or the tonic component, which represent the baseline level of skin conductivity and (2) the skin conductance response (SCR), which represents phasic increases in the amplitude of skin conductivity. These deflections are often a result of a psychophysiological response to discrete environmental stimuli, though spontaneous deflections which are not stimuli-related are common for most people as well. The Ledalab computer program (Benedek and Kaernbach 2010), a free MATLAB-based software for the analysis of raw EDA data was utilized to calculate the following five EDA indexes:

- nSCR: the number of significant phasic SCRs within a chosen time window. Based on a trial and error procedure a threshold value of 0.1μS (microsiemens) was used to distinguish between significant and non-significant responses in outdoor environments.
- AmpSum: the sum in μS of the significant SCRs within the chosen time window
- PhasicMax: the local amplitude of the largest SCR deflection in μS within a time window
- GlobalMean: the average skin conductivity level within the chosen time window
- MaxDeflection: the maximum level of skin conductivity within this window

The first three indexes take into account the magnitude of local deflections. Therefore, these indicators are expected to be more useful in detecting momentary changes in outdoor environments. On the other hand the last two indexes, GlobalMean and MaxDeflection are global in their nature, meaning that they take into account absolute values of EDA and overlook the local amplitude of SCRs. They are expected to be less useful in out-of-the-lab studies, in which environmental conditions are not controlled and the absolute EDA levels may change rapidly regardless of mental state (due to increased heat leading to sweating, for example).

## Methods

The Wearables

Two commercial off-the-shelf wearables were tested, Empatica's E4 wristband and MS Band (see Figure 1). These bands were chosen due to their large number of sensors and the simplicity of installation, which allowed easy implementation for participants in everyday conditions. To the best of our knowledge, these were the only two devices to offer such characteristics in the time when the study took place. Both bands are designed to be worn on the wrist and they include a comparable set of sensors (Table 1). The physiological signals that the bands can record include (maximum temporal resolution of the data is given in parentheses where 1 Hz equals to one sample every one second): HR (E4: 1 Hz; MS: 1 Hz), interbeat intervals, EDA (E4: 4 Hz; MS: 0.2 Hz), skin temperature (E4: 1 Hz; MS: 0.04 Hz), and blood volume pulse (E4: 64 Hz; MS: n/a). Both bands rely on photoplethysmography technology to extract cardiovascular signals; they also include a three-axis accelerometer. According to Microsoft's official manual, it should be noted, the EDA sensor is meant to detect whether the band is worn on the wrist and not to perform accurate EDA measurements. Additional information that can be recorded with the MS Band includes distance travelled, elevation, number of steps, and environmental data (i.e., ambient temperature, atmospheric pressure, and brightness). In contrast to the E4 band, the MS Band is equipped with a built-in GPS, though the raw GPS data cannot be extracted. In the current study we utilized the HR, interbeat intervals (that are used to extract HRV) and EDA physiological signals to detect changes in mental state.

*--Insert Figure 1 about here--*

**Figure 1**: Empatica E4 band (top) and Microsoft's MS Band 2 (bottom)

The E4 band was designed to record physiological signals for research and clinical purposes. As such, it includes a convenient interface through which data can be uploaded to a secure cloud storage in both streaming and offline modes. The MS Band, on the other hand, does not claim to supply clinically tested measurements. It is marketed as a smart band that allows the wearer to monitor fitness and healthy lifestyle on a daily basis. The band does not permit straightforward raw data exportation. In this study, we used a third-party Android mobile application called Data Log for Microsoft Band to log the band's measurements. Table 1 presents the technical specifications and performances of both bands in greater detail.

*--Insert Table 1 about here--*

Procedure and Participants

A homogenous sample of 15 male students (mean age: 21.8 years; standard deviation: 1.74 years) was recruited. Participants received a €25 voucher as an incentive. Participants were asked to give informed and written consent before the beginning of the experiment. The research protocol was approved by the Ethics Review Board of the Faculty of Social and Behavioural Sciences at Utrecht University (FETC17-086). Three participants were excluded from the final sample due to missing data.

The experiment included a walk along a predefined route in the city center of Utrecht, the Netherlands. The route was designed to include a variety of urban landscapes ranging from green spaces to a walk along a main road (see route and segments in Figure 2). Participants arrived independently at the meeting point at Utrecht's central train station. The E4 and MS Band were affixed to their wrists, the E4 on the dominant hand and the MS Band on the other hand. In addition, participants were also equipped with a GPS logger (BT-Q1000XT) that tracked their location every second. Participants were instructed to follow a research assistant on a 3-kilometer-long walk while keeping a distance of 20–30 meters. This strategy was applied for a number of reasons. First, it allowed participants to focus only on the walk while avoiding distractions such as reading a map. Second, it guaranteed that all participants took the same route and walked at a similar speed. In order to stimulate a stressful situation participants had to cross a main road without a traffic light. For reasons of safety, participants were informed about the crossing in advance and the crossing itself was controlled by the trained research assistant, who walked side by side with the participant at that point. At the end of the walk participants were asked to complete a questionnaire in which they were asked to rank their subjective walking experience in each segment from 1 – most relaxing to 8 – most stressful based on a map of the trail.

Data Processing and Analysis

Using an overlay operation within GIS environment, each GPS sample of each participant was assigned with the pre-defined characteristics of the walking route segments (see Figure 2). At the second phase physiological signals recorded by the bio-sensors were matched with the GPS information based on the timestamps of the datasets. Since GPS samples were recorded in a 1Hz rate (i.e. once every second) and EDA information of the E4 band was recoded in a 4Hz rate (i.e. 4 samples every second), the mean EDA level of each four consecutive samples was assigned to a GPS reading. HRV and several EDA indexes which are aggregative in their nature and cannot be calculated for each GPS position were calculated per geographic segment. They were than compared using a t-test to detect significant differences in physiological reactions. In particular, the analysis focused on the stressful crossing episode (a 30-second time window with a 5-second offset starting from the point at which participants began the crossing) that was compared with signals recorded a few minutes earlier in one of the more neutral, less stressful environments (segment 9, see Figure 2). In this episode EDA indexes—nSCR, AmpSum, PhasicMax, GlobalMean, and MaxDeflection as well as HRV—SDNN, pNN50, LF/HF—indexes were utilized (for more details, see section 'Physiological Signals as Markers for Mental States').

**Results**

Descriptive Statistics of Geo-referenced Physiological Signals

Figure 2 presents the walking route divided into segments and Table 2 summarizes the mean level of the physiological signals recorded by the E4 band, EDA level and HR and the subjective ranking scores of all participants for each walking segments. The table reveals that on average participants had a steady increase in EDA levels along the walk (a similar trend was reported during outdoor measurements in Osborne and Jones 2017). This is most likely a result of the increase in body temperature and sweating during the walking activity. Absolute EDA levels of one participant are represented by the color of the GPS sample points at the small figure in the bottom. This figure shows the main fields of data that each GPS sample was assigned with. EDA level for this participants ranged between 0.060μS (recorded in the central station) to 2.931μS (during the bus ride).

*--Insert Figure 2 about here—*

**Figure 2**: The walking route divided into segments. The EDA levels along the route of one of the participants (small figure).

*--Insert Table 2 about here--*

Detecting the Impact of Stressful Urban Episodes on Mental States

**EDA**: Due to the low sampling rate (0.2Hz) and poor performance of the MS Band in measuring EDA (data hardly showed any variation), only the results of the E4 are presented here. We could not find significant differences that will indicate that a momentary change in mental state occurred when passing through the different types of environments except for the case of the crossing segment. Table 3 presents the results of five computed indexes extracted from the raw EDA data for the stressful crossing and compare them with a more neutral walking environment. The number of significant SCRs (nSCR index) shows that participants had 12.4 significant SCRs under neutral conditions and 17.87 SCRs during the more stressful street crossing. A paired *t*-test was employed to detect differences between the baseline (i.e. neutral) condition and stressful one (i.e. road crossing). In accordance with the hypothesis, the stressful condition resulted in significantly higher nSCR index ($t=-2.777$, $p=0.015$) indicating on an increased activation of the sympathetic nervous system. The table also shows that for 10 out of 15 participants (67 percent) the nSCR index was higher during the stressful crossing than it was in the more neutral setting; for these individuals, the physiological response was in line with our expectations. Similarly, the sum in μS of the significant local SCRs (AmpSum index) and the maximum local SCR amplitude within the response window (PhasicMax) also supported the hypothesized increase in EDA in response to a stressful situation. Though significant and in the expected direction, the results of both GlobalMean and MaxDeflection should be carefully interpreted. Since these indexes represent the absolute EDA levels, they might reflect the gradual increase in skin conductivity that participants experienced along the walk. Such an increase might have occurred due to the physical effort and environmental conditions that participants encountered (i.e., increased sweating and humidity) and not necessarily as a result of a psychophysiological reaction to a stressful event.

--*Insert Table 3 about here*--

**HR and HRV**: With both the E4 and the MS Band, no significant difference was detected in HR level between the walking segments including for the stressful walking segments. In the case of HRV, the E4 band generated incomplete interbeat interval datasets during the walks, and thus did not allow for the calculation of HRV indexes. We therefore used only data from the MS Band to compute HRV indexes. Similarly to the EDA indexes, our analysis revealed no significant differences between the different walking segments except for the case of the stressful crossing episode. In this segment a statistically significant difference (paired *t*-test, $t=-2.459$, $p=0.028$) between the neutral conditions and the stressful ones was found in the frequency domain measurement LF/HF index (see 'Physiological Signals as Markers for Mental States' section above). This finding indicates that momentary stressful situations evoke physiological cardiovascular reactions (HRV indicators) which could potentially be detected through wearables.

## Discussion

The study demonstrated how the continuous stream of geo-referenced physiological signals can be contextualized and enriched using location tracking technologies. This technique allows characterizing the surrounding environment as well as some aspects of the activity one is encountering (e.g. walking, crossing a road, entering a shop, using public transportation). While biosensors are now becoming a popular tool for the daily monitoring of physical activity (El-Amrawy et al. 2015; Li et al. 2017; Wright et al. 2017), extracting meaningful information related to mental dimensions of behavior using these sensors seems somewhat more complex. Nonetheless, our findings seem to be in line with other similar studies that indicated that though limited and inferior to lab equipment, off-the-shelf wearables can produce meaningful documentation of physiological signals when enriched by spatial context that is recorded by location technologies such as GPS and subjective assessments of types of spaces (El-Amrawy et al. 2015; Cormack et al. 2016; Osborne and Jones 2017). More specifically, we found EDA measurements of the E4 to be useful in detecting stressful episodes in less controlled outdoor conditions. Though less conclusive, cardiovascular signals were also found to be useful markers for monitoring the change in mental state during the stressful crossing. Indicators such as HRV, it should be remembered, might be more ambiguous in cases in which signals are recorded for short periods of less than five minutes (Healey and Picard 2005; Appelhans and Luecken 2006), as was the case in our study. Moreover, even though we had a relatively small sample of participants, some of the results did support the feasibility of utilizing heart indicators in naturally occurring environments using existing wearables.

While we could detect changes in mental state during the road crossing, an important question still remains; why did the exposure to other environments that commonly known to have therapeutic qualities (e.g. green and blue spaces) did not result in changes in mental state. From a technical-methodological perspective, it might be that the devices are not sensitive enough to detect such changes. This may require the implementation of more sensitive devices or of a larger sample. Similarly, it could be that the specific changes in mental state that are evoked by green and blue environments are not reflected in EDA and cardiovascular indicators. In this case, other physiological signals and corresponding sensors should be employed (see for example Aspinall et al. 2015 who implemented electroencephalography). It could also be the case that the environmental exposure in the study (a brief walk through green, blue and urban environment) did not generate any therapeutic or adverse effect on mental state. The attention restoration theory (Kaplan and Kaplan 1989), for example, attributes cognitive restoration qualities to natural environments, but in case a person is not cognitively overloaded, it might be that this person will not experience changes in mental state. The theory also suggests that in order to demonstrate therapeutic outcomes the environment should include specific characteristics (e.g. "soft fascinations") which might have been absence from the environments in the study. Future studies should therefore test the devices in different environments and/or for longer exposure times. Finally, it should be noted that the literature regarding the beneficial qualities of green and blue spaces is often ambiguous as to the actual impact of these environments on our mental state (see for example: Bowler et al. 2010; Gascon et al. 2015). Biosensing techniques may help shed some light on this ambiguity.

While the fact that some of the results were found significant and in accordance with expectations is promising, it is important to note several limitations of the present study. First, our sample was relatively small and homogeneous and future studies should include larger, more diverse samples in terms of gender, age and socio-economic background. Nevertheless, since the focus of the study was methodological and since there is no reason to assume a methodological bias between different groups of the population, the results are expected to be useful for other groups as well. Second, the outdoor measurements were taken for short periods of time and under highly controlled conditions. Although this made the interpretation of the results easier, it raises questions as to whether more natural and "noisy" measurements could be similarly interpreted (Osborne and Jones 2017). The implementation of this technique under "real-world" uncontrolled conditions for long periods of time will make the real challenge of this method.

While much has been learned about the analysis of physiological signals in lab experiments, best practices for the utilization of such measurements in naturally occurring environments is limited (some exceptions include: Hartig et al. 2003; Healey and Picard 2005; Osborne and Jones 2017; Shoval, Schvimer, and Tamir 2018a). In particular, it is essential that researchers find ways to detect meaningful psychophysiological reactions and to correctly pair them with the evoking stimuli (Bakker, Pechenizkiy, and Sidorova 2011). The need to find valid methods that eliminate potential cofounders is also closely related to this issue. The latter is especially crucial in the case of stimuli-rich outdoor environments and when measurements are conducted during physical activity. In order to achieve this, researchers must collect rich contextual information about the activity and the physical and social environments with which the participants are engaging continuously (Osborne and Jones 2017). In our study, we utilized GPS information and geographical layers to better understand the environmental context of the situation. Implementing activity diaries and utilizing additional complementary data collection tools such as smartphones (Birenboim et al. 2015; Birenboim and Shoval 2016; Eskes et al. 2016) and various other sensors (Sagl, Resch, and Blaschke 2015) may be required in less controlled settings. Such information should allow researchers to control and eliminate potential cofounders and to reach more reliable interpretations of the results garnered.

Future studies should take advantage of the growing availability of detailed geographical information to further enrich the environmental characterization and tempo-spatial resolution of analysis. For example, each GPS location can be assigned with relevant data such as the density level of the buildings within a specified radius, the number of trees and green elements in sight, number of food and commercial outlets, pollution levels, crowd (based on cellular information), weather and more rather than simply relying on predefined categories as was done here. However, as noted above, the theoretical and practical limitations of this approach should be acknowledged.

## Conclusions

With the advances in wearable technology and increased public awareness about healthy lifestyles, it seems likely that in the near future we will witness a surge in new commercial devices and complementary software (Blaauw et al. 2016) both for more popular self-monitoring and for clinical usage. This study demonstrated that the potential of monitoring mental states in real-world conditions using wearables exists—but much work has yet to be done before such devices can be utilized in standard research or clinical procedures. From a technological point of view, the reliability of wearables in measuring relevant physiological signals during daily activity should still be improved. Due to the numerous applications that could utilize such technology, including the monitoring of physical and mental wellbeing, there is a strong commercial incentive for manufacturers to develop such technology.

Finally, it is crucial to ascertain that ethical and societal aspects related to sensing techniques are being properly addressed. Privacy is obviously of high concern when it comes to e-health in general and sensing technologies more specifically. The field raises techno-ethical questions regarding data ownership, storage and management as well as legal concern regarding proper usage (Nissenbaum and Patterson 2016). Other ethical concerns revolve around the appropriate implementation of the technology. The utilization of the technology to discipline workers through wellness initiatives (Moore and Piwek 2017) is only one example in which the technology may lead to dystopian outcomes. Therefore, it is important that the expected technological development will be accompanied by social and ethical research efforts regarding the impact of technology adoption on human behavior and desirable societal usage (Schüll 2016; Moore and Piwek 2017).

**Acknowledgments:** The research was supported by the interdisciplinary Healthy Urban Living research program of Utrecht University. Marco Helbich was partly funded from the European Research Council (ERC) under the European Union's Horizon2020 research and innovation program (grant agreement No 714993).

**Additional information**

**Author information**

**Amit Birenboim**
AMIT BIRENBOIM is a senior lecturer at the Department of Geography and the Human Environment, Tel Aviv University, Zelig 10, 6997801, Tel Aviv, Israel. E-mail: abirenboim@tauex.tau.ac.il. His research interests include the implementation of sensors and advanced location tracking technologies to the study of spatial behavior, health geographies and the study of individuals' wellbeing and momentary experiences in urban environments.

**Martin Dijst**
MARTIN DIJST has been appointed as full professor of Urban Development and Spatial Mobility at Utrecht University, the Netherlands in 2009. In December 1, 2017, he has been appointed as director of the department of Urban Development and Mobility at LISER, Porte des Sciences 11, 4366, Esch-sur-Alzette, Luxembourg. His research interests focus on activity and travel behaviour, exposures to (un)healthy environments, social interactions with people and urban metabolism.

**Floortje E. Scheepers**
FLOORTJE E. SCHEEPERS is a psychiatrist, professor of innovation in mental health and head of the Department of Psychiatry at the University Medical Center in Utrecht. PA 85500, Utrecht, the Netherlands E-mail: f.e.scheepers-2@umcutrecht.nl. Her research focuses on innovation, applied big data statistics and e-health in mental health care.

**Maartje Poelman**
MAARTJE POELMAN is an Assistant Professor in Public Health Sciences in the Department of Human Geography and Spatial Planning, Utrecht University, Princetonlaan 8a, 3584 CB, Utrecht, the Netherlands. E-mail: m.p.poelman@uu.nl. Her research interests focus on the food environment-diet relationship, incorporating the influence of individual determinants like mental stress, emotions and daily activities.

**Marco Helbich**
MARCO HELBICH is an Associate Professor in the Department of Human Geography and Spatial Planning, Utrecht University, Princetonlaan 8a, 3584 CB, Utrecht, The Netherlands. E-mail: m.helbich@uu.nl. His research interests focus on geocomputational techniques and spatiotemporal analytics to address human-environment relations in cities.

**Table 1**: Technical specifications and performance comparison: The E4 vs. the MS Band

| | Criteria | E4 | MS Band 2 |
|---|---|---|---|
| **Specifications** | Sensors/data | **Physiological signals**: HR (1Hz), interbeat intervals, EDA (4Hz), skin temperature (1Hz), blood volume pulse (64Hz)<br>**Spatial/environmental**: 3-axis accelerometer<br>**Other**: Event mark button to manually tag events | **Physiological signals**: HR (1Hz), interbeat intervals, EDA (0.2Hz), skin temperature (0.04Hz)<br>**Spatial/environmental**: 3-axis accelerometer, GPS (raw data not accessible), ambient temperature, atmospheric pressure (barometer), brightness |
| | User interface | A convenient interface for uploading logged data to secure cloud storage. Live data streaming through mobile devices is available. | Third-party applications are required to log raw data.<br>* In order to maintain the integrity of the data, participants are required to stay within a short distance of the recording smartphone at all times |
| | Other technical specifications (published by manufacturer) | Battery (continuous sampling): 20h+ streaming, 36h+ logger mode<br>Charging: <2h<br>Storage: 60h of raw data can be stored on the band (internal memory). Includes a streaming mode in which storage is not limited.<br>Water resistant (to splashes) | Battery (continuous sampling): n/a<br>Charging: <2h<br>Storage: Raw data cannot be stored on the internal memory of the band. Data is stored in the memory available on the smartphone and is dependent on it<br>Water resistant (to splashes) |
| | Other features | API that allows the development of own applications (Android, iOS) | API that allows the development of own applications (Android, iOS, windows phone) |
| | Price | High price tag of ~US$1600 | Relatively low price tag ~US$250 makes it an affordable device. |
| **Performance** | EDA | Our tests showed that EDA information under both lab and real-world conditions were useful. | EDA information was found unsuitable for detecting changes in mental states. |
| | HR / HRV | Static posture: high-quality interbeat interval data. However, data series is incomplete; many missing values.<br>Walking: Incomplete data during walking activity; insufficient for the extraction of HRV indexes. | Acceptable quality of heart signal information during both static and walking measurements. |

For a detailed explanation about HR (heartrate), interbeat intervals (which are used to extract heartrate variability indexes) and EDA (electrodermal activity) see section ' Physiological Signals as Markers for Mental States' above.

**Table 2**: Mean scores of HR, EDA and subjective ranking of stress level of each walking segment (table)

| # | Environment type | EDA | HR | Subjective Ranking* |
|---|---|---|---|---|
| 1 | Central station (indoor) | 2.212 | 96.0 | 6.73 |
| 2 | Busy junction | 2.831 | 104.4 | 6.33 |
| 3 | Neighborhood commercial street (Lombok) | 3.443 | 101.3 | 5.27 |
| 4 | Neighborhood street (Lombok) | 3.888 | 103.1 | |
| 5 | Blue space 1 (canal) | 4.262 | 102.7 | 2.73 |
| 6 | Blue space 2 | 4.724 | 103.3 | |
| 7 | Green space (urban park) | 4.757 | 103.5 | 1.20 |
| 8 | non-commercial street 2 | 5.445 | 105.1 | 3.47 |
| 9 | Pedestrians street | 5.804 | 102.9 | |
| 10 | Walk along a main road | 6.254 | 102.2 | 5.27 |
| 11 | Road crossing | 6.625 | 103.1 | |
| 12 | Walk to bus station | 7.135 | 97.4 | |
| 13 | Bus ride | 6.992 | 84.5 | 5.00 |

*Subjective rankings of the walking segments attractiveness made by the participants. Lower numbers indicate higher ranking of attractiveness. Some segments were clustered in the questionnaire.

**Table 3**: A comparison between EDA indexes measured during participants' outdoor walks in a neutral setting and in a more stressful situation using the E4 band.

| | Neutral setting (mean) | Stressful crossing (mean) | Paired *t*-test | Percent of participants with expected response |
|---|---|---|---|---|
| nSCR | 12.40 | 17.87 | *t*=2.777 *p*=0.015 | 67% |
| AmpSum (µS) | 3.58 | 10.85 | *t*=2.764 *p*=0.015 | 93% |
| PhasicMax (µS) | 1.65 | 3.56 | *t*=3.828 *p*=0.002 | 87% |
| GlobalMean (µS) | 6.35 | 7.18 | *t*=2.190 *p*=0.046 | 60% |
| MaxDeflection (µS) | 0.56 | 1.26 | *t*=2.464 *p*=0.027 | 100% |